\documentclass[12pt,a4paper,english]{article}
\usepackage[authoryear]{natbib}
\usepackage{graphicx}
\usepackage{isorot}
\usepackage{geometry}
\newcommand{\degr}{$^{\circ}$}  
\newcommand{\ignore}[1]{}
\newcommand{\ex}{$\times $\,} 

\makeatother
\begin{document}

\title{{A review of wildland fire spread modelling, 1990-present}\\
      { 2: Empirical and quasi-empirical models}}

\author{A.L. Sullivan}

\maketitle
{\small

\begin{center}

Ensis\footnote{A CSIRO/Scion Joint Venture} Bushfire Research \footnote{current address: Department of Theoretical Physics,\\
Research School of Physical Sciences and Engineering\\
The Australian National University, Canberra 0200, Australia. }\\
PO Box E4008, Kingston, ACT 2604, Australia

email: Andrew.Sullivan@ensisjv.com or als105@rsphysse.anu.edu.au

phone: +61 2 6125 1693, fax: +61 2 6125 4676

\end{center}}

\begin{center}version 3.0 \end{center}

\setcounter{secnumdepth}{0}
\setlength\parskip{\bigskipamount}
\setlength\parindent{0pt}

\begin{abstract}
In recent years, advances in computational power and spatial data analysis
(GIS, remote sensing, etc) have led to an increase in attempts to model the
spread and behaviour of wildland fires across the landscape. This series of
review papers endeavours to critically and comprehensively review all types of
surface fire spread models developed since 1990. This paper reviews models of an
empirical or quasi-empirical nature. These models are based solely on the
statistical analysis of experimentally obtained data with or without some
physical framework for the basis of the relations. Other papers in the series
review models of a physical or quasi-physical nature, and mathematical
analogues and simulation models. The main relations of empirical models are
that of wind speed and fuel moisture content with rate of forward spread.
Comparisons are made of the different functional relationships selected by
various authors for these variables.
\end{abstract}


\section{Introduction}

\subsection{History}

An empirical model is one that is based upon observation and experiment and not
on theory.  Empiricism has formed the basis for much of the scientific and
technological advances in recent centuries and generally provides the benchmark
against which theory is tested. The study of fire and combustion in general was
mainly an empirical endeavour, directed primarily toward application of
combustion to industrial processes (for example, the industrial revolution of
the late 1700s-1800s), until early in the previous century when the physical or
theoretical approach had matured to the point of providing significant advances
in understanding and prediction.  The development of physical understanding of
other forms of combustion (i.e. unintentional or uncontrolled fire) in general
and wildland fires in particular, did not occur, however, until only very
recently (in the last few decades) \citep{Sullivan2007b}.

While there had always been a great general interest in unintentional fire in
urban settings \citep{Williams1982}, for instance, the Great Fire of London in
1666, or the Chicago Fire on October 1871--prevention, control,
prediction--unintentional fire in wildlands received much less attention,
mainly due to the relatively little impact such fires have on the general
populace. The study of the behaviour of fires in wildland regions has
traditionally been driven by the needs of those practitioners involved in
wildland resource management--foresters for the most part--for whom
understanding this natural phenomenon was critical to the success of their
work.

Despite the fact that practically no region of the world (except for
Antarctica) is free from such fires, much of the work in this field was
galvanised in the United States following the devastating 1910 fires in the
mid-west \citep{Pyne2001}, where workers such as \citet{Hawley1926} and
\citet{Gisborne1927,Gisborne1929} pioneered the notion that understanding of
the phenomenon of wildland fire and the prediction of the danger posed by a
fire could be gained through measurement and observation and theoretical
considerations of the factors that might influence such fires.
\citet{Curry1938,Curry1940}, and \citet{Fons1946} brought a rigorous physical
approach to the measurement and modelling of the behaviour of wildland fires
that set the benchmark for wildland fire research for decades following.

In addition to the work conducted in the US, through the Federal US Forest
Service and State agencies, other countries became increasingly involved in
wildland fire research, primarily through their forest services--the Canadian
Forest Service, the Commonwealth Forestry and Timber Bureau (later absorbed
into the Commonwealth Scientific and Industrial Research Organisation (CSIRO)
in conjunction with various state authorities in Australia--although many other
countries such as South Africa, Spain, Russia, France, Portugal to name a few,
have also had significant impact on wildland fire research.

Since the early 1990s, European Union countries have committed significant
funds towards wildland fire research, resulting in a boom period for this
research in mainly Mediterranean countries and a major shift in focus away from
the pioneering three (US, Canada and Australia).

During the past two decades, the direction of much of the wildland fire
research has been toward the use of fire as a resource management tool in the
form of hazard reduction burning or the study of ecological effects of fire
(e.g. \citet{Gill1981, Goldhammer1990, Abbott2003}.

\subsection{Empirical modelling}

The focus of empirical modelling of wildland fire in the past has been on the
determination of the key characteristics used to describe the behaviour of the
fire. These generally have been the rate of forward spread (ROS) of the head
fire (that portion of the fire perimeter being blown downwind and normally of
much greater intensity that the rest of the fire perimeter), the height of the
flames, the angle of the flames, and the depth of flames at the head, although
other characteristics such as rate of perimeter or area increase may also be of
some interest.

While observations of wildfires or fires lit intentionally for other purposes
(such as hazard reduction or prescribed fires) have been used in the
development of empirical models of fire behaviour, the predominant method has
been the lighting of `experimental' fires--fires whose only purpose is that of
an experimental nature. This method can be divided into four parts.  Firstly,
the characterisation and quantification of the fuel and terrain in which the
fire will lit (the slowly varying variables, which has included fuel load, fuel
height, moisture content, bulk density, combustion characteristics, slope,
etc.). Secondly, the observation and measurement of the atmospheric environment
(the quickly varying variables, wind speed and direction, air temperature,
relative humidity, etc.). Thirdly, the lighting, observation and measurement of
the fire itself (its speed, spread, flame geometry, combustion rate, combustion
residues, smoke, etc.). Fourthly, the statistical correlation between any and
all of the measured quantities in order to produce the model of fire behaviour.
Many workers have chosen to limit or control the possible natural variation in
many quantities by conducting experimental fires in laboratory conditions which
aids in the analysis of such fires.

The primary use of such models has been to estimate the likely spread in the
direction of the wind (and potential for danger to firefighter safety) for
suppression planning purposes, much of which has traditionally been conducted
in the form of simple `back of the envelope' calculations for plotting on a
wall map. Due to this simple need, empirical fire spread models have
traditionally been one dimensional models in which the independent variable
that is predicted is the rate of forward spread of the head of the fire in the
direction of the wind.  The rather pragmatic nature of these models, their
relatively straightforward implementation, their direct relation to the
behaviour of real fires, and, perhaps most importantly, their development by
for the most part by forestry agencies for their own immediate use, have meant
that empirical fire spread models have gained acceptance with wildland fire
authorities around the world and to varying degrees form the basis for all
operational fire behaviour models in use today.

\subsection{Operational models}
In the United States, the quasi-empirical model of \citet{Rothermel1972} forms
the basis of the National Fire Danger Rating System
\citep{Deeming1977,Burgan1988} and the fire behaviour prediction tool BEHAVE
\citep{Andrews1986}. This model is based on a heat balance model first proposed
by \citet{Fransden1971} and utilised data obtained from wind tunnel experiments
in artificial fuel beds of varying characteristics and from Australian field
experiments of grassfires in a range of wind speed conditions to correlate fire
behaviour with measured input variables. The model of \citet{Rothermel1972} and
associated systems have been introduced to a number of countries, particularly
Mediterranean Europe.

In Australia, the predominant operational fire spread prediction systems have
been the McArthur Grassland \citep{McArthur1965,McArthur1966} and Forest
\citep{McArthur1967} Fire Danger Rating Systems (FDRS), and the Forest Fire
Behaviour Tables for Western Australia (commonly called the Red Book)
\citep{Sneeuwjagt1985}, based on the work of \citet{Peet1965}.  Both McArthur's
systems and the Red Book are purely empirical correlations of observed fire
behaviour and measured fuel and environmental variables from mainly field
experimental fires augmented by well-documented wildfires . More recently, the
CSIRO Grassland Fire Spread Meter (GSFM) \citep{CSIRO1997, Cheney1997b} based
on the empirical modelling of \citet{Cheney1998} has replaced the McArthur
Grassland FDRS as the preferred tool for predicting fire behaviour in
grasslands. This, too, is based on field experimentation and documented
wildfire observations.

In Canada, the quasi-empirical Fire Behaviour Prediction (FBP) System
\citep{FCFDG1992} forms part of the Canadian Forest Fire Danger Rating System
(CFFDRS) and is the culmination of 60 years of research effort in fuel moisture
and fire behaviour \citep{VanWagner1998, Taylor2006}.  Almost 500 fires were
used in the construction of the FBP system, of which approximately 400 were
field experiments, the remainder well-documented observations of prescribed and
wild fires. The CFFDRS has been introduced and implemented in a number of
countries, including New Zealand, Mexico and several countries of south-east
Asia.

The main characteristic of all but the CSIRO GSFM is that these systems were
based primarily on small ($<$1 ha) experimental or laboratory fires and
augmented with wildfire observations\footnote{Another interesting
characteristic is that the CSIRO GFSM model was the only one published in a
peer-reviewed journal; all the others were published as technical reports by
the associated organisations.}. The series of experiments upon which the CSIRO
GFSM was based \citep{Cheney1993} was the first to use experimental burning
plots of which the smallest was 1 ha (See review of this model below).

\subsection{Background}

This series of review papers endeavours to comprehensively and critically
review the extensive range of modelling work that has been conducted in recent
years. The range of methods that have been undertaken over the years represents
a continuous spectrum of possible modelling \citep{Karplus1977a}, ranging from
the purely physical (those that are based on fundamental understanding of the
physics and chemistry involved in the behaviour of a wildland fire) through to
the purely empirical (those that have been based on phenomenological
description or statistical regression of fire behaviour). In between is a
continuous meld of approaches from one end of the spectrum or the other.
\citet{Weber1991a}, in his comprehensive review of physical wildland fire
modelling, proposed a system by which models were described as physical,
empirical or statistical, depending on whether they accounted for different modes
of heat transfer, made no distinction between different heat transfer modes, or
involved no physics at all. \citet{Pastor2003} proposed model descriptions of
theoretical, empirical and semi-empirical, again depending on whether the model
was based on purely physical understanding, of a statistical nature with no
physical understanding, or a combination. \citet{Grishin1997} divided
models into two classes, deterministic or stochastic-statistical. However,
these schemes are rather limited given the combination of possible approaches,
and, given that describing a model as semi-empirical or semi-physical is a
`glass half-full or half-empty' subjective issue, a more comprehensive and
complete convention was required.

Thus, this review series is divided into three broad categories: Physical and
quasi-physical models; Empirical and quasi-empirical models; and Simulation and
Mathematical analogous models. In this context, a physical model is one that
attempts to represent both the physics and chemistry of fire spread; a
quasi-physical model attempts to represent only the physics. An empirical model
is one that contains no physical basis at all (generally only statistical in
nature), a quasi-empirical model is one that uses some form of physical
framework upon which to base the statistical modelling chosen. Empirical models
are further subdivided into field-based and laboratory-based. Simulation models
are those that implement the preceding types of models in a simulation rather
than modelling context. Mathematical analogous models are those that utilise a
mathematical precept rather than a physical one for the modelling of the spread
of wildland fire.

Since 1990 there has been rapid development in the field of spatial data
analysis, e.g. geographic information systems and remote sensing. Following
this, and the fact that there has not been a comprehensive critical review of
fire behaviour modelling since \citet{Weber1991a}, I have limited this review
to works published since 1990. However, as much of the work that will be
discussed derives or continues from work carried out prior to 1990, such work
will be included much less comprehensively in order to provide context.

\subsection{Previous reviews}

Many of the reviews that have been published in recent years have been for
audiences other than wildland fire researchers and conducted by people without
an established background in the field. Indeed, many of the reviews read like
purchase notes by people shopping around for the best fire spread model to
implement in their part of the world for their particular purpose. Recent
reviews (e.g. \citet{Perry1998,Pastor2003}; etc), while endeavouring to be
comprehensive, have offered only superficial and cursory inspections of the
models presented. \citet{Morvan2004a} take a different line by analysing a
much broader spectrum of models in some detail and conclude that no single
approach is going to be suitable for all purposes.

While the recent reviews provide an overview of the models and approaches that
have been undertaken around the world, mention must be made of significant
reviews published much earlier that discussed the processes in wildland fire
propagation themselves. Foremost is the work of \citet{Williams1982} which
comprehensively covers the phenomenology of both wildland and urban fire, the
physics and chemistry of combustion, and is recommended reading for the
beginner. The earlier work of \citet{Emmons1963,Emmons1966} and \citet{Lee1972}
provides a sound background on the advances made during the post-war era.
\citet{Grishin1997} provides an extensive review of the work conducted in
Russia in the 1970s, 80s and 90s. \citet{Chandler1983} and \citet{Pyne1996}
provide a useful review of the forestry approach to wildland fire research,
understanding and practice.

The first paper in this series discussed those models based upon the
fundamental principles of the physics and chemistry of wildland fire behaviour.
This particular paper will discuss those models based directly upon only
statistical analysis of fire behaviour observations or models that utilise some
form of physical framework upon which the statistical analysis of observations
have been based.  In this paper, particular distinction is made between
observations of the behaviour of fires in the strictly controlled and
artificial conditions of the laboratory and those observed in the field under
more naturally occurring conditions.

The last paper in the series will focus upon models concerned only with the
simulation of fire spread over the landscape and models that utilise
mathematical conceits analogous to fire spread but which have no real-world
connection to fire.

\section{Empirical models}

The following sections identify and discuss those empirical and quasi-empirical
surface-only fire spread models that appeared in the literature since 1990. It
is interesting to note the observation of \citet{Catchpole2000} that the
majority of new models that have been developed in recent years have been the
result of efforts to initially develop and validate local fuel models required
for the implementation of the BEHAVE (based on Rothermel) fire behaviour
prediction system. Many researchers obviously felt that it was far easier to
start from scratch with a purpose built model than to try to retrofit their
local conditions into an existing model.  Table \ref{Table:SummEmp} summarises
the empirical models discussed in this review.

Due to the varied nature of the empirical models presented here, including the
fuels and weather conditions under which the data for the construction of the
models were collected, the size and number of experimental fires and purposes
for which the models were developed, it is difficult to compare them side by
side. One possible method is the relationship between rate of forward spread
(ROS) and wind speed. Wind speed is widely accepted as being the dominant
variable determining the forward speed of a fire front.  The reasons for this
are cause for significant debate, ranging from the reduction in angle of
separation of flame to unburnt fuel to increased turbulent mixing of
combustants.  Regardless of the mechanics of the process, the empirical
approach to modelling fire spread must cater for this process and is manifested
in the functional form chosen to represent it. Fuel moisture content (FMC) is
also a key variable in determining rate of spread and this is also discussed.
Fire spread models for fuel layers other than surface fuels, such as crown
fires or ground fires, are not covered.

\subsection{Canadian Forest Service (CFS) - Acceleration (1991)}
While the Canadian Forest Service (CFS-accel) work \citep{McAlpine1991a} is not
a model of fire spread as such, it does address a major concern of fire spread,
namely the acceleration in rate of fire spread from initiation.  The assumption
is that a fire will attain an equilibrium rate of spread for the prevailing
conditions (the prediction of which is the primary aim of all fire spread
prediction systems discussed here). The form of the function for the time to
reach this equilibrium ROS is assumed to be exponential based on models
proposed by \citet{Cheney1981} and \citet{vanWagner1985} (as cited by
\citet{McAlpine1991a}).

29 experimental fires were conducted in a wind tunnel with a fuel bed 6.15 m
long by 0.915 m wide consisting of \emph{Pinus ponderosa} needles or
excelsior\footnote{Excelsior is wood shavings cut into long thin strands} of
varying fuel load and bulk density.  Four wind speeds (0, 0.44, 1.33 and 2.22 m
s$^{-1}$) measured at mid-flame height were used. Temperature and relative
humidity were held constant at 26.7\degr C and 80\%.  Equilibrium ROS was
assumed to occur after 2.0 m of forward spread and determined using linear
regression of averaged fire location and time measurements.

Acceleration was modelled as an allometric (power law) function asymptoting to
the equilibrium ROS with two coefficients, one based on the equilibrium ROS
(which eliminated differences in fuel properties and integrated all other
burning condition variables) and the other on the wind speed.  The model was
found to well represent the laboratory data but observations of elapsed time to
equilibrium ROS did not coincide with point source field observations of other
authors which were much greater and also dependent on wind speed.

\subsection{CALM Spinifex (1991)}

The Western Australia Department of Conservation and Land Management (CALM)
Spinifex model \citep{Burrows1991} was developed from 41 experimental fires
conducted in predominantly spinifex (\emph{Triodia basedowii} and
\emph{Plectrachne schinzii}) fuels on relatively flat sand plains.  These fires
were lit using drip torches to create lines of fire up to 200 m long
perpendicular to the wind direction. Fuel particle dimension and arrangement
were measured for individual clumps; fuel distribution, quantity and moisture
content were measured using line transect methods.  Bare ground between clumps
was also measured.  Wind speed and direction, air temperature and relative
humidity were measured at 10-min intervals.  Wind speed ranged over 1.11 - 10 m
s$^{-1}$ and FMC over 12-31\%. Fire spread was measured using metal markers
placed near the flame front at intervals of 1-4 mins and later surveyed.  Fires
were allowed to spread until they self-extinguished. The range of ROS was
0-1.53 m s$^{-1}$. Data gathered were analysed using multiple linear regression
techniques.

\citet{Burrows1991} found that above a threshold wind speed zone (3.33-4.72 m
s$^{-1}$), in which flames are tilted sufficiently to bridge the gap between
hummocks \citep{Bradstock1993, Gill1995}, the ROS varied with the square of the
wind speed (R$^2$ = 0.85).  Below the threshold wind speed zone, which depends
on the percent cover of fuel (ratio of percentage of area covered by hummocks
to bare ground), the fire does not spread.  The higher the percent cover, the
lower the threshold wind speed required. A lesser, negative, linear correlation
was determined with FMC. Percent cover and air temperature were also found to
influence the ROS but much less than either wind or FMC. Fuel load and other
fuel characteristics were found not to be important.

\subsection{Canadian Forest Fire Behaviour Prediction (CFBP) System (1992)}

The Canadian Forest Fire Behaviour Prediction (CFBP) System is a component of
the Canadian Forest Fire Danger Rating System (CFFDRS) \citep{Stocks1991},
which also incorporates the Canadian Forest Fire Weather Index (CFWI) System.
The CFFDRS is the result of continuing research into forest fire behaviour
since the mid-1920s and has undergone several incarnations in that time.  The
current CFFDRS system came into being in the late 1960s in the form of a
modular structure.  The first major component to be completed was the CFWI in
1971, which provided a relative measure of fuel moisture and fire behaviour
potential for a standard fuel type, and has been revised several times since
its introduction \citep{VanWagner1987}. While there have been several interim
editions of the CFBP, the first of which appeared in 1984 \citep{Lawson1985},
it was not until 1992 that a final version of the prediction system was
released \citep{FCFDG1992, Taylor2006} and thus is covered in this review.

The CFBP system, following on from the long-established Canadian approach to
studying wildland fire, is based on the combined observations of nearly 500
experimental, prescribed and wild fires in 16 discrete fuel types covering 5
major groups: coniferous, deciduous, mixed wood, slash and grass fuels. The
experimental work on which the system is based was conducted by individual
researchers working in specific fuel types and locales across the country using
a variety of methods and published in a variety of places (initially including
the 1966 work of McArthur in Australian grasslands, later replaced by the data
of \citet{Cheney1993} \citep{FCFDG1992}). \citet{Alexander1991} provides an
overview of the methods used since the 1960s to obtain the dataset from the
CFBP was derived, but which, due to a number of factors (including
technological improvements) evolved over the years. The result is a system
constructed by a small group of dedicated researchers over a period of 20 years
that has broad applicability to a wide range of fuels and climates.

Experimental burn plots varied in size from 0.1 ha up to 3 ha
\citep{Alexander1991}, with the majority being less than 1 ha.  Ignition
methods included both point ignitions as well as line-ignitions.  Wind speed
unaffected by the fire was measured at 10-m in the open (or converted a 10-m in
the open equivalent). Experiments were usually conducted in the late afternoon
in order to attain maximum burning conditions for the day.  ROS was normally
measured by visual observations of fire passage over predetermined distances.
For point ignition experiments, metal tags were placed at the head and flanks
of the fire and surveyed afterwards.

The final version of the CFBP system works in conjunction with the CFWI system
to determine an Initial Spread Index (ISI) for the standard fuel type (pine
forest) and based solely on fine FMC and wind speed. The functions chosen for
the effect of wind speed and fine FMC on the ISI are exponential (exponent
0.05039) for wind, and a complicated mix of exponential and power law
(exponents -0.1386 and 5.31 respectively) for FMC \citep{VanWagner1987}. No
quantification of performance of these functions is given.

To predict ROS, the ISI is modified by a Build-up Index (BUI), which is a
fuel-specific fuel consumption factor that includes fuel moisture. Predicted
ROS is the headfire ROS on level terrain under equilibrium conditions, thereby
implicitly including effects of acceleration and crowning \citep{FCFDG1992}.
The effect of slope \citep{VanWagner1977a} and crown fire transition effects
\citep{VanWagner1977b} then modify the basic ROS. Recent work of the
International Crown Fire Modelling Experiment \citep{Stocks2004} has
investigated the behaviour of fully-developed crown fires (which is not covered
in this review as it is outside the scope of surface fire spread).

\subsection{Button (1995)}

\citet{Marsden1995b} presented a model for the prediction of ROS and flame
height of fires in Tasmanian buttongrass moorlands, described as largely
treeless communities dominated by sedges and low heaths \citep{Marsden1995a}.
The behaviour of 64 fires (of which 44 were experimental fires, 4 test fires,
11 fuel reduction fires and 5 wildfires) at 12 sites was measured. Experimental
burns were conducted on blocks of either 0.25 or 1.0 ha with ignition line
lengths of 50 or 100 m respectively under a limited range of weather
conditions.  ROS was measured by either using metal tags thrown at different
times or by timing the passage of flames past pre-measured locations. For
experimental fires, wind speed and direction, temperature and relative humidity
were measured at 10 m, and wind speed only at 1.7 m above ground level, all
averaged over 1-3 min periods. Meteorological data for non-experimental fires
were collected using handheld sensors at 1.7 m. Data ranged from 0.19 - 10 m
s$^{-1}$ for wind speed and 8.2-96\% for FMC, and 0 - 0.92 m s$^{-1}$ for ROS.

\citet{Marsden1995b} found surface wind speed, dead FMC and fuel age (time
since last fire) to be the key variables affecting ROS, with wind being the
dominant factor. Age and FMC each accounted for 15 to 20\% of the observed
variation in ROS.  A power law with an exponent of 1.312 was used to describe
the effect of wind, whereas both the FMC and fuel age were modelled as
exponential functions (FMC decreasing, age increasing to a maximum at about 40
years). Rates of spread of the back and flank of the fires were found to be
approximately 10\% and 40\% of the head ROS, respectively.

\subsection{CALM Mallee (1997)}

\citet{McCaw1997} conducted a large-scale field experiment in \emph{Eucalyptus
tetragona} mallee-heath community in south-west Western Australia. Shrubs $<$
1.0 m tall comprised more than half the plant species present. Burn plots 200 m
\ex 200 m were established in 20-year-old fuel in flat terrain.  A
semi-permanent meteorological site was set up 500 m from the experimental plots
recording 30 min averages of temperature and relative humidity at 1.5 m and
wind speed and direction at 2 m. During each experiment, mean wind speed and
direction at a location up to 250 m upwind of the plot were measured at heights
of 2 m and 10 m at 30 s intervals. FMC was measured using 5 samples of four
fuel components (3 dead and 1 live) collected post-fire within 30 min of
ignition. Wind speed at 10 m in open ranged from 1.5-6.9 m s$^{-1}$, FMC
4-32\%. Experimental fires were ignited using a vehicle-mounted flame thrower
to establish a line perpendicular to the prevailing wind up to 200 m long. Fire
spread was measured using buried electronic timers (placed on a 24-point grid)
equipped with a fusible link that melted on exposure to flames.  ROS ranged
0.13-0.68 m s$^{-1}$.

Isopleths representing the position of the fire front at successive time
intervals were fitted to the grid of timer data for each plot using a
contouring routine based on a distance-weighted least squares algorithm.  ROS
up to 0.67 m/s and fireline intensities up to 14 MW/m were recorded.  Fires
were found to spread freely when the FMC of the dead shallow litter layer
beneath the low shrubs was $<$ 8\%.  Forward ROS was modelled as a function of
the wind speed in the open at 2 m and FMC of the deep litter layer.  These
accounted for 84\% of the variation in ROS.  A power function (exponent 1.05)
and an exponential (coefficient -0.11) were chosen to describe wind (measured
at 2 m) and FMC influences respectively \citep[page 142]{McCaw1997}. Good
agreement between the model and observations of rate of spread of a limited
number of prescribed and wild fires (up to ROS = 1.1 m/s), although observed
ROS of a wildfire burning under extreme fire danger conditions was
over-predicted by 30\%.

\subsection{CSIRO Grass (1997)}

The CSIRO Grassland Fire Spread Meter \citep{CSIRO1997} is a cardboard circular
slide rule that encapsulates the algorithms developed by \citet{Cheney1998} for
fire spread in natural, grazed and eaten-out grassland pastures.  These
algorithms are based primarily on the results of experiments conducted in
annual grasses of the Northern Territory with the aim of determining the
relative importance of fuel characteristics on rate of forward spread of large
unconstrained fires, particularly fuel load \citep{Cheney1993}, augmented by
large experimental fires conducted in open woodland \citep{Cheney1995} and
detailed observations of 20 wildfires. 121 experimental fires were carried out
on a flood plain in a range of fuel treatments under a variety of weather
conditions \citep{Cheney1993,Cheney1995} in prepared blocks ranging in size
from 100 $\times$ 100 m to 200 $\times$ 300 m. These fires were predominantly
lit from lines ranging in length from 30 to 175 m, although there were also a
number of point ignitions, and allowed to burn freely.  The range of fuel
treatments included mowing and removing cuttings, mowing and retaining
cuttings, or leaving the grass in its natural state. Two distinct grass species
(\emph{Eriachne burkittii} and \emph{Themeda australis}), of different height,
bulk density and fineness were present.

Fuel characteristics (height, load, bulk density, etc.) were measured on four
transects through each plot approximately every 25 m. In addition to remote
standard 10 m and 2 m meteorological stations, the wind speed at 2 m was
measured at the corner of each plot and averaged for each ROS interval. FMC
samples were taken before and after each fire. ROS and flame depth were
measured from a series of rectified time-stamped oblique aerial photographs of
each fire. Wind speed ranged from 2.9 - 7.1 m s$^{-1}$, FMC 2.7-12.1\%, and ROS
0.29-2.07 m s$^{-1}$.

\citet{Cheney1995} found the growth of the fires to be related to wind speed
and the width of the head fire normal to the wind direction.  They found that
the width of the fire required to achieve the potential quasi-steady ROS for
the prevailing conditions increased with increasing wind speed, and the time to
reach this quasi-steady ROS was highly variable.  ROS was found to depend on
the initial growth of the fire, the pasture type (natural, grazed or
eaten-out),  wind speed and live and dead FMC.  Utilising the notion of
potential quasi-steady ROS and a minimum threshold wind speed for continuous
forward spread, \citet{Cheney1998} developed a model of fire spread assuming a
width necessary to reach the potential ROS. This model uses wind speed, dead
FMC and degree of curing to predict the potential (i.e. unrestricted) ROS for
the prevailing conditions.  Above a threshold of 5 km h$^{-1}$ the ROS is
assumed to have a power function (with an exponent less than 1 (0.844))
relation with the wind speed.  This wind speed function is similar to that
proposed by \citet{Thomas1961}, in which a power function with exponent of just
less than 1 was found. Less than the threshold, the ROS is linear with wind
speed and dominated by dead FMC.

\subsection{Heath (1998)}
A cooperative research effort from a number of Australasian organisations,
Heath \citep{Catchpole1998b} utilises observations of 133 fires (comprising a
mix of experimental (95), prescribed (22) and wild (16) fires) conducted in
mixed heathland (heath and shrub) fuels.  This includes 48 experiments
conducted by \citet{Marsden1995b} in buttongrass.  Only experimental and
prescribed fires were used in model development; wildfire observations were
used for validation.

In mixed heathland (comprising heath, scrub and gorse in New Zealand and mixed
species including Banksia, Hakea and Allocasuarina in Australia), fuel age
ranged from 5-25 years.  Fires were lit as lines of unstated length on slopes
$<$ 5\degr. Due to the disparate nature of the researchers involved, methods
for measuring variables varied from experiment to experiment.  Wind speed was
generally measured by handheld anemometry at 2 m at 20 s intervals and averaged
over the life of the fire. Wind speed ranged from 0.11-10.1 m s$^{-1}$ and ROS
0.01-1.00 m s$^{-1}$. Fuel load does not appear to have been measured but fuel
height was. FMC was measured in some cases and modelled in others using
pre-established functions based on air temperature and relative humidity.

Wind speed was found to account for 53\% of the variation in ROS. Aerial dead
fuel (i.e. those fuels not in contact with the ground) FMC was found not to be
significant. Fuel height was highly significant and with wind accounted for
70\% of the variation in ROS.  A power function of wind speed (exponent 1.21)
was used to describe this variation.  A power function was also used for fuel
height (exponent 0.54).

The model was found to perform reasonably well for the selection of wildfires,
considering the paucity of available data and necessary assumptions about the
involved fuel characteristics (fuel height, moisture etc.) but could be
improved with more variables. The wind power function does fail for zero wind
but was found to better fit the data than an exponential growth function.

\subsection{PortShrub (2001)}

\citet{Fernandes2001,Fernandes1998} presented a model developed from field
experiments and observations of prescribed burns conducted in four different
types of shrub in flat terrain of Portugal. He found that \citet{Rothermel1972}
did not predict observed ROS well. 29 fires were conducted on flat ($<$3\degr
 slope) in gorse, low heath, tall heath and tall heath/tree mix. Fine aerial
live and dead FMC was sampled prior to each burn. Meteorological variables
(wind speed, air temperature, relative humidity) were measured at 2 m in the
open using either a fixed weather station placed near the burn plot or upwind
with handheld instruments. Fires were lit as lines of length 10 m in
experimental fires and 100 m in prescribed burns. ROS was measured by recording
time of arrival of the head fire at reference locations. Wind speed ranged
0.28-7.5 m s$^{-1}$, FMC 10-40\% and ROS 0.01-0.33 m s$^{-1}$.

ROS was significantly correlated with wind speed (1\% level) and less so with
RH, temperature, and aerial dead FMC (5\%).  Other fuel characteristics were
also found to affect ROS but were strongly intercorrelated and thus could not
be separated, however, preference was given to fuel height. The initial model
found a power law (exponent 1.034) for wind speed. However, as the model
predicted no ROS in zero wind, an exponential function (coefficient 0.092) was
subsequently incorporated. The final model, with an exponential decay function
for dead FMC (coefficient -0.067) and power function (exponent 0.932) for fuel
height, improved the overall performance of the model (R$^2$ = 0.91). The model
was also found to predict well the data sets of other authors and be in close
agreement with other field studies (e.g.
\citet{Marsden1995b,Cheney1998,Catchpole1998b}).

\subsection{CALM Jarrah I (1999)}

\citet{Burrows1999a,Burrows1994} conducted a series of 144 laboratory
experiments (54 wind-driven, 6 no wind, 13 backing, 34 with slope, 15 point
ignition) using fallen leaves and twigs ($<$6 mm) placed on a 4 m long by 2 m
wide table set in a large shed. Wind was supplied by four domestic fans
calibrated to give a desired wind speed over the fuel bed. FMC was varied by
uncontrolled ambient conditions and wetting prior to burning. It ranged from 3
to 14\%. Fires were lit along the 2 m upwind edge using cotton wick soaked in
methylated spirits and allowed to burn for 50 cm before measurements commenced.
ROS was measured by recording time taken to reach end of fuel bed. Wind was
varied from 0.0 to 2.1 m s$^{-1}$ with mean 1.06 m s$^{-1}$. ROS ranged
0.002-0.075 m s$^{-1}$.

In wind-driven fires, no relationship between fuel load and forward ROS was
found. Most variation in ROS was due to wind speed (correlation coefficient
0.94). ROS was negatively related to FMC (correlation coefficient -0.31).
Backing ROS was found to be directly related to fuel load.

At wind speeds $<$ 0.83 m s$^{-1}$, ROS was relatively insensitive to wind.
Above this value, ROS was found to vary linearly with wind speed.  However, a
power function (exponent 2.22) was used to model wind speed effect on ROS.  An
inverse linear function was used for FMC.  This model was found to underspecify
ROS $>$ 3.33 m s$^{-1}$ with an error variance that increased with ROS.

\subsection{CALM Jarrah II (1999)}

\citet{Burrows1999b, Burrows1994} studied four series of fire behaviour data
obtained from field experiments and fuel reduction burns on flat to gently
sloping terrain in Jarrah (\emph{Eucalyptus marginata}) forest in south-west
Western Australia to test Jarrah I and other models for forest fire spread.
Fuel was characterised by a layer of dead leaves, twigs, bark and floral parts
on the forest floor with low ($<$0.5 m, 30\% cover) understorey of live and
suspended vegetation. Plots were 100 m wide \ex 200 m long. 56 of 66 total
plots were lit from lines of 50-100 m length, the remainder being point
ignitions.

Historical (pre-fire) weather data (including rainfall, temperature, relative
humidity, wind speed and direction at 2 hourly intervals) were obtained from
nearby permanent weather stations. During each experiment a portable weather
station approximately 50 m from the fire recorded wind speed at 1.5 m and 10 m
and temperature and relative humidity at 1.5 m at 5 minute averages. FMC was
measured at the time of ignition. Wind speed at 10 m in the open ranged
0.72-3.33 m s$^{-1}$ and FMC 3-18.6\%.

ROS was measured by recording the time of arrival at a grid of predetermined
locations, along with other fire characteristics, after first allowing the fire
to spread 20 - 40 m ($\simeq$ 15 min) in order for it to attain a quasi-steady
ROS for the prevailing conditions. The position of the flames in relation to
the grid was mapped at 5 min intervals. ROS ranged 0.003-0.28 m s$^{-1}$.

Unlike the laboratory findings \citep{Burrows1999a}, Burrows here found a
non-linear relation between wind speed and ROS. A power function (exponent
2.674) was selected.  FMC was determined to also be a power function (exponent
-1.495).  Like the laboratory findings, fuel load was not found to correlate
with ROS.  The model was found to underpredict ROS of large wildfires burning
under severe conditions.

\subsection{Gorse (2002)}
\citet{Baeza2002} conducted field experiments during spring and autumn in gorse
shrublands of eastern Spain with the aim of developing a prescribed burning
guide. Fuels were 3, 9 and 12 years old and were replicated 3 times resulting
in a total of 9 fires.  Plots were 33 m \ex 33 m and were burnt under low ($<$
1.39 m s$^{-1}$ at 2 m) wind, utilising headfire spread for the 3-year-old fuel
and backing fires for the other two age classes. Meteorological data was
recorded at 2 m at 15 min intervals.  Fuel characteristics were recorded along
5 parallel transects 5 m in length. FMC was measured from 10 samples of the
most abundant species collected prior to ignition and ranged from 22-85\%,
presumably including both live and dead fuels\footnote{It should be noted that
the authors dried their fuels at 80\degr for 24 hours which is much less than
the generally accepted 104\degr  C for 24 hours (e.g. \citet{Cheney1993}),
perhaps resulting in lower than actual FMC values--see discussion on
Measurement issues.}. Ignition technique is not specified. ROS was measured by
recording the time to travel a fixed distance within the plot and ranged
0.004-0.039 m s$^{-1}$.

It was found that FMC was the dominant factor affecting ROS in a linear manner
(coefficient 0.487).  The combination of heading and backing propagation
negated any consistent effect of wind speed on ROS.

\subsection{PortPinas (2002)}

\citet{Fernandes2002} developed a model for the behaviour of fires in maritime
pine (\emph{Pinus pinaster} stands in northern Portugal under a range of fire
weather conditions that occur outside the wildfire season for the purpose of
improving the understanding of prescribed fire for hazard reduction.  Six study
sites in mountainous terrain with forests founded by plantation or regeneration
following fire events and aged 14 to 41 years were established. Fuel complexes
were dominated by litter, shrubs or non-woody understorey (e.g. grass) types.
Extensive destructive and non-destructive sampling to quantify the fuels was
undertaken along transects in each experimental plot. Four strata of fine fuel
layers were defined: shrubs, herbs and ferns, surface litter and upper duff.
Experimental plots were square, 10-15 m wide, and defined by 0.3 to 1.2 wide
control strips assisted by a hose line during burning.

Wind speed was measured continuous at 1.7 m above ground approximately 10 m
from each experimental plot. Three composite fuel moisture samples (one litter,
one duff and one live) were sampled at random locations prior to ignition.

94 experimental fires for fire behaviour studies were conducted when slope and
wind direction were aligned within 20\degr. Line ignition occurred 2 m from the
windward edge to allow both forward and backing spread observations. Fire
behaviour measurements used 1.5-m-high poles located at regular distances along
the plot axis as reference points. ROS was determined by recording the time at
which the base of the fire front reached each pole. Flame height and flame
angle were estimated visually and used to calculate flame length.  Wind speed
ranged from 0.3 - 6.4 m s$^{-1}$,  surface dead FMC ranged 8 - 56\%, air
temperature 2 - 22\degr C and relative humidity 26 -96\%.  ROS ranged 0.004 -
0.231 m s$^{-1}$.

\citet{Fernandes2002} found that three existing models underestimated ROS with
significant differences between predicted and observed values, as much as
8-fold in one case. Undertaken non-linear least=squares analysis, they found
that slope and wind speed were the most significant variables with dead FMC in
a less significant role. A power law function with wind speed only (exponent
0.803) explained 45\% of the variation in ROS.  If wind speeds less than 0.83 m
s$^{-1}$ were excluded, the correlation coefficient increased to 0.996.  Slope
alone explained 30\% of the variation.  The final model selected for
litter-shrub fuels (the general case for maritime pine stands) involved wind
speed (power law, exponent 0.868) dead surface FMC (exponential, coefficient
-0.035), slope and understorey fuel height.  Fuel height was selected as could
be considered a surrogate for the overall fuel complex structure effect on ROS.
The model was then adapted through changes in constants to predict ROS in
litter and non-woody understorey complexes. No assessment of the performance of
this model was reported.

\subsection{Maquis (2003)}

\citet{Bigili2003} conducted a series of 25 field experiments in open, level
shrubland of maquis fuel in southwestern Turkey.  Average height of the fuel
was 0.53 m and fires were conducted under a range of wind and fuel conditions.
Each fire plot was 20 m wide by 30 m long. A meteorological station recorded
air temperature, relative humidity, wind speed and precipitation at 1.8 m
daily. Fuel characteristics were measured from random destructive sampling
prior to the experiment series. Live and dead FMC was sampled immediately prior
to ignition. During each fire, wind speed, temperature and relative humidity at
1.8 m were recorded at 1 min intervals using the automatic meteorological
station. These were averaged over the period of fire spread. Wind speed ranged
0.02-0.25 m s$^{-1}$, FMC 15.3-27.7\%.

Fires were lit with a drip torch along the upwind (20 m long) edge to quickly
establish a line fire and were allowed to propagate with the wind across the
plot. ROS was measured by recording time of arrival at a series of
predetermined locations and ranged 0.01-0.15 m s$^{-1}$. ROS was strongly
correlated with wind speed; a linear function explained 71\% of observed
variation. FMC was found not to have any significant effect on ROS, attributed
to the narrow range studied. The final model used a linear function of wind
speed (coefficient 0.495) and total fuel load, with an R$^2$ = 0.845.

\section{Quasi-empirical models}

Where the data gathered from experimental observation is analysed using a
physical framework for the functional relationships between dependent and
independent variables, a quasi-empirical model results.  The degree to which
the physical framework controls the structure of the model can vary but the
nature of the model is essentially based upon the observed data (which
differentiates it from quasi-physical models which use data solely for
parameterisation).  Table \ref{Table:SummQE} summarises the quasi-empirical
models discussed below.

\subsection{TRW (1991)}

\citet{Wolff1991} presented the results of laboratory experiments conducted in
a purpose-built wind tunnel 1.1 m wide by 7 m long with a moveable ceiling. The
fuel layer was vertical match splints (1.3-4.4 mm in diameter) set in a ceramic
substrate. Wind speed varied from 0-4.7 m s$^{-1}$, ROS ranged from 0-0.007 m
s$^{-1}$. The results confirmed the theoretical treatment conducted by
\citet{Carrier1991}, in which it was hypothesised that the dominant heat
transfer mechanism in such a set-up would be a mix of convection and diffusion
(i.e. `confusion') heating that would result in a relationship in which the ROS
would vary as the square root of the wind speed normalised by the fuel load. If
radiation was the predominant preheating mechanism, it was hypothesised that
the variation would be as the power of 1.5 rather than 0.5.

Wolff \emph{et al.} found that not only did the width of the fuel bed play an
important part in determining the ROS but also the total width of the wind
tunnel itself. The narrower the fuel bed, and the facility, the slower the ROS.
It was suggested that a narrower fuel bed forced air away from the fuel bed due
to drag considerations in the fuel. A series of experiments with tapering fuel
beds and working section confirmed this. If the fuel bed and working section
was too narrow, ROS ceased.

\subsection{NBRU (1993)}

\citet{Beer1993, Beer1991} investigated the interaction of wind and fire spread
utilising a series of 18 small-scale wind tunnel (length 40 cm, height 16 cm)
experiments using a single row of match splints in wind ranging from 0.0 to 9 m
s$^{-1}$. ROS ranged from 0.004-0.38 m s$^{-1}$. Rather than a single
continuous function to describe the relationship between wind speed and ROS,
Beer put forward the hypothesis that there exists a critical characteristic
(threshold) wind speed that affects ROS with different wind speed functions
above and below this value. Below the threshold Beer found a normalised (by the
threshold wind speed) power function (exponent 0.5).  Above the threshold, Beer
found a normalised power function (exponent 3.0).  Beer postulates that the
choice of the value is related to the wind speed at which the wind shear is
strong enough to generate flame billows and that this value corresponds to a
mid-flame wind speed of 2.5 m s$^{-1}$. Above this value it is thought that the
flames remain within the fuel bed rather than above it.  Beer attempted to fit
this model to observations of grassfire behaviour but could not.

\citet{Beer1995} further explored the effects of wind on fire spread through
simplified (match splints) fuel.  His extension of a simple geometric model of
fire spread in no wind to include wind (based on geometry of wind-tilted flame
and distance between fuel elements), in which the ROS-wind function is a
complicated solution to a set of equations to determine the critical time for
flame immersion of adjacent fuel elements, did not perform well.  This was
attributed to assumptions about the characteristics of the flame and a constant
ignition temperature.  Beer concludes that a single simple power law or
exponential is unlikely to be a correct mathematical description for the
ROS-wind speed relation.

\subsection{USFS (1998)}

\citet{Catchpole1998a} conducted an extensive series of
environmentally-controlled wind tunnel experiments and used the results, in
conjunction with energy transfer considerations, to develop a spread model,
USFS (United States Forest Service). 357 experimental fires were carried out on
a fuel bed 8 m long by 1 m wide in a 12-m long wind tunnel of 3 m square cross
section. Four fuels with different surface-area-to-volume ratios (two sizes of
poplar excelsior, ponderosa pine needles and ponderosa pine sticks) were chosen
to be reasonable approximations to natural fuel layers. Temperature and
relative humidity were controlled to produce a range of FMC, 2\% to 33\%
(although the majority of fires were carried out at ambient values of 27\degr C
and 20\% RH giving an FMC range of 5-9\%).  Wind speed above the tunnel's
boundary layer ranged from 0.0 to 3.1 m s$^{-1}$. Rate of spread was measured
at 0.5 m intervals using photovoltaic diodes placed 25 mm above the fuel bed to
record the time of arrival of the flame front.

Utilising the conservation of energy model of \citet{Fransden1971}, Fransden's
\citeyearpar{Fransden1973} effective heating number, a propagating flux model
that is linear in packing ratio, and an exponential decay function for FMC, the
authors built a model of fire spread very similar in its construction to that
of \citet{Rothermel1972} except that they used the heat of ignition of a unit
mass of fuel (which comprises the heat of pyrolysis and heat of dessication)
rather than the heat of pre-ignition as used by Rothermel. A power function for
wind was then fitted to the data and an exponent of 0.91 determined. Although a
cubic polynomial function was found to better fit the data, the authors chose
the power function as it was more consistent with data from wildfire
observations.

\subsection{Coimbra (2002)}

\citet{Viegas2002} presents a quasi-empirical model of fire spread that
utilises the geometry of the fire perimeter to determine the forward spread
rate.  The main conceit of this notion, previously proposed in
\citet{Viegas1998a} and \citet{Viegas1998b}, is that a line fire lit at an
angle to a slope or wind gradient undergoes a translation and rotation of the
fireline in order to spread with the maximum rate in the direction of the
gradient. Extensive laboratory experimentation utilising a double-axis tiltable
fuel bed (1.6 m \ex 1.6 m) for a range of forward/back and left/right slopes
was used to develop the model. The fuel was \emph{Pinus pinaster} needles with
an FMC determined by ambient conditions (ranging 10\% - 15\%). 23 experimental
fires were conducted, with 10 fires of varying slope and 13 fires of varying
inclination. \citet{Viegas2002} found a maximal rotation velocity at an
inclination angle of 60\degr\, but was unable to convert this to a forward ROS.
However, Viegas does develop a fire perimeter propagation algorithm in which
the perimeter is treated as a continuous entity that will endeavor to align
itself with the gradient, through this proposed rotation mechanism, to an angle
of approximately 60\degr. The translation and rotation hypothesis, however,
ignores a basic observation of the evolution of flanking spread and instead
assumes that spread at non-parallel angles to the slope or wind gradient must
be driven by a headfire.

\citet{Viegas2005} attempts to extend these ideas to describe the phenomenon of
`fire blow-up' based on the concepts of fire `feedback effects'. Viegas
proposes the existence of a positive dynamic feedback between the ROS of a fire
and the flow velocity driving the fire such that the fire accelerates
exponentially. He uses some of the results of experimental fires burnt in a
``canyon'', a doubly-sloped tray, in no wind and a range of canyon slopes and
inclinations to parameterise his model and the remainder to test it.  Viegas
treats all data for all slope and inclination combinations as independent and
continuous. As a result, his model increases ROS exponentially, resulting in
extremely rapid acceleration--what he describes as blow-up.  However,
categorised by slope, rather than treated continuously, the ROS data actually
asymptotes to a reasonable number in each case, which in most cases confirms
the long-held rule of thumb of doubling the flat ground ROS for every 10
degrees increase in slope \citep{McArthur1967, VanWagner1988}.
\citet{Viegas2006} conducts a parameteric study of this model and determines
that fires in light and porous fuels are more likely to exhibit `eruptive'
behaviour than fires in heavy and compacted fuels.

Viegas's extrapolation of this model to fatal wildfire incidents is tenuous at
best and really only proves the widely accepted acceleration up a slope. Other,
more robust, theories of unexpected fire behaviour resulting in fatalities are
probably more applicable (e.g. \citet{Cheney2001}).

\subsection{Nelson (2002)}

\citet{Nelson2002} extended the quasi-empirical work of \citet{Nelson1988}
utilising the laboratory data of \citet{Weise1997} to build a trigonometric
model of fire spread that combines wind and slope effects into a single
combined `effective' wind speed.

The \citet{Nelson1988} model utilised the dimensional analysis of fire
behaviour of \citet{Byram1966}, where three dimensionally homogeneous (i.e.
dimensionless) relations were derived: 1) the square root of the Froude number,
2) a buoyancy number relating convective heat output to rate of buoyancy
production, and 3) the ratio of combustion time to time characteristic of flame
dynamics. \citet{Nelson1988} then used spread observations from 59 experimental
fires (a total of 44 lab and 21 field, some deleted) and mixed and matched the
dimensionless relations until they found a combination that gave a reasonable
correlation. They derived a dimensionless form of ROS and wind speed, which,
when fitted to the data and converted back to dimensions, gave a power law
relation between wind speed and ROS (exponential 1.51). As the maximum wind
speed used to obtain the data was 3.66 m s$^{-1}$ and maximum observed ROS was
0.271 m s$^{-1}$, \citet{Nelson1988} acknowledged the need for higher wind
speed experiments. ROS was also found to be a function of fuel load (power
function, exponent 0.25) and residence time (inversely proportional). FMC was
considered to be accounted for in the estimate of fuel load and residence time.

Rather than the traditional approach used by \cite{McAlpine1991b} where the
equivalent wind speed for a slope-only ROS was determined, \citet{Nelson2002}
used the concept of vertical buoyant velocity and the slope angle component of
this wind to construct a slope-induced wind which was then added vectorally to
the ambient wind across slope. Nelson extended the dimensional analysis of
\citet{Nelson1988} to then determine ROS. The resultant equations, which do not
apply to flanking or backing fires due to the assumption about convective
heating through the Froude number, were then compared against the data of
\citet{Weise1997}, gathered from 65 experiments in a portable tilting wind
tunnel using vertical paper birch sticks as the fuel bed in a variety of wind
and slope configurations, ranging from 0.0 to 1.1 m s$^{-1}$ and -30\degr to
+30\degr. The effective wind speed was found to correlate linearly with ROS.

\section{Discussion}

\subsection{Wind speed function}

As stated earlier, one method of comparing the structure of each of the above
models is to examine the form of the functional relationship between ROS and
wind speed chosen by the authors.  Table \ref{Table:WindFn} summarises the
models discussed and the form of the wind speed function chosen.  Also listed
are the experimental bounds of the wind speed and ROS.

\begin{figure}[ph!]
  \centering
  \includegraphics[width=16cm]{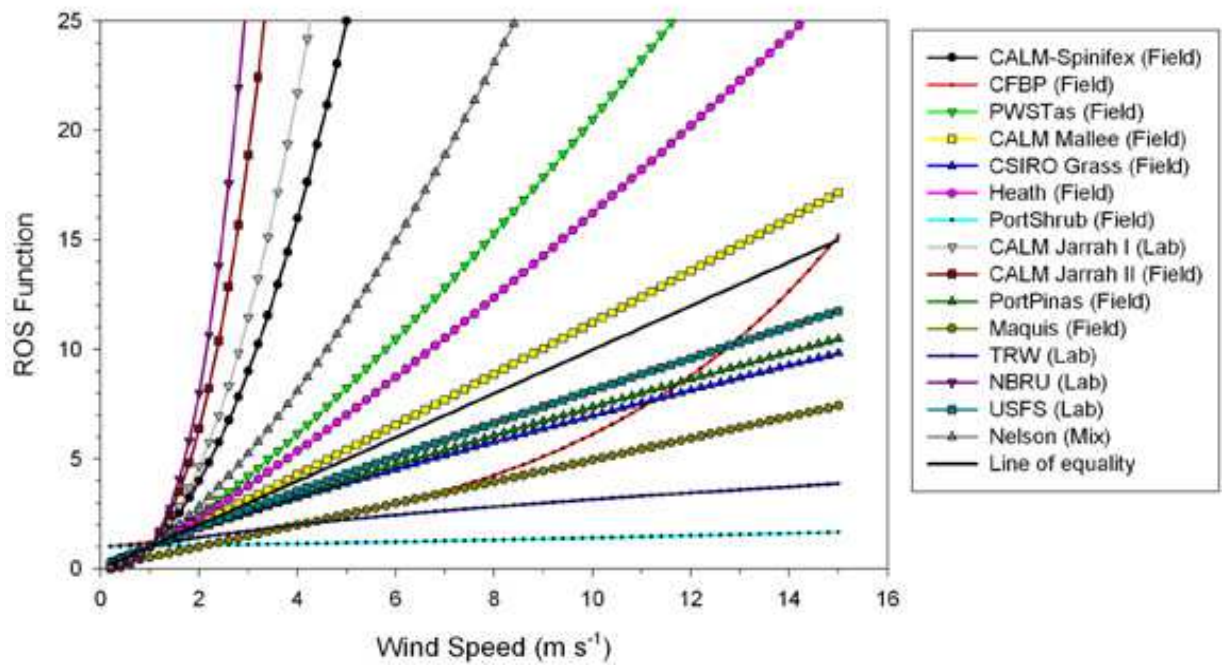}\\
  \caption{Graph of functional relationships between wind speed and rate of
  forward spread used in various empirical and quasi-empirical fire spread models.
  The relationship is only indicative of effect on ROS as the full model in each case
  may also include effects of other variables such as fuel moisture content as well as
  increases or decreases in wind speed due to measurement at different heights.}\label{Fig:WindFn}
\end{figure}

Only three of the models for which the wind function is given are not power
functions, PortShrub and CFBP are exponential and Maquis is linear.  Of the
remaining models, 3 models have exponents less than one: TRW, CSIRO Grass and
USFS.  The remaining wind functions all result in non-linear increases (Figure
\ref{Fig:WindFn}) in the ROS that will result in a speed greater than the wind
speed driving it, which is unphysical \citep{Beer1991}.  (While this is also
the case also for CFBP as illustrated here for wind speeds $>$ 15 m s$^{-1}$,
the ROS function is further modified in the CFFDRS system by fuel-specific
functions which can reduce the predicted ROS below the wind speed.) The reason
for this choice of function appears to be the desire by the modellers to fit
data at low wind speeds (including zero). Many of the models had ranges of wind
speed that were fairly low ($<$ 3 m s$^{-1}$).

The few models that were based on large ranges of wind speed in field
experiments (with the exception of CALM Jarrah II) tended to result in power
functions with exponents less than one. CSIRO Grass is the highest power
function less than one and is very similar in form to the linear function of
Maquis over the given range. \cite{Fendell2001}, in their brief review of the
topic including a number of older models, found that the wind power function
exponents ranged from 0.42 \citep{Thomas1967} to 2.67 \citep{Burrows1999b}.

There seem to be two key factors in the choice of functional relationship used
to describe fire spread and wind speed.  The first is the need to fit the
function through the origin.  In many cases, particularly laboratory
experiments, zero wind speed is taken as the default state and thus any
continuous function must not only fit the data of non-zero wind, but also of
zero wind.  This is discussed in greater detail below. The second is that for
the most part, the range of wind speeds studied (again particularly in the
laboratory) is very small. As can be seen in Figure \ref{Fig:WindFn}, any
function, be it cubic or very shallowly linear, performs rather similarly at
low wind speeds ($<$1.5 m s$^{-1}$).

It is interesting to note that in the full range of functions presented here,
the nearly median wind function in Figure \ref{Fig:WindFn} (i.e. Heath) is the
result of the combination of multiple datasets, experimental methods and
authors, perhaps resulting in a middle ground of approaches.  Many physical
models of fire spread (e.g. \citet{Grishin1984a, Linn1997}) have observed
linear functional relationships between wind speed and ROS, suggesting that
power law functions with exponents close to unity may have a more fundamental
basis.

In their validation of the performance in Mediterranean shrub fuels of seven
wildland fire spread models, including the CFBP and Rothermel,
\citet{Sauvag2001} found that a model's performance is not related to the
model's complexity and that even the most simple model (in this particular case
a local fire officer's rule of thumb based on a linear discount of the wind
speed) performed as well as more complicated models such as CFBP.

\subsection{Threshold wind speed}\label{Threshold}

One important aspect differentiating the various choices of wind function, is
the ideal of a continuous function that includes zero wind speed.  It has been
noted previously \citep{Burrows1991} that fires in discontinuous fuels such as
spinifex have a minimum wind speed required before forward spread is achieved.
This notion was extended further by \citet{Cheney1998} to define a threshold
wind speed at which fires spread forward \emph{consistently}.  The argument is
that fires burning in low winds in the open respond to eddy fluctuations in the
wind flow (resulting in near circular perimeter spread after a long period) and
do not spread in a continuous consistent manner until the wind speed exceeds a
certain threshold. Above this threshold, the fire spreads forward in a manner
directly related to the wind speed.

The choice of threshold value is dependent then upon the method of measuring
the wind speed (location, height, period, etc) and the fuel type in which the
fire is burning (taller fuels reduce the wind speed reaching the fire).
\citet{Cheney1998} chose a 1.39 m s$^{-1}$ open wind speed threshold for open
grasslands.  \citet{Fernandes2002} found that wind speed explained more
variation in ROS when wind speeds below 0.83 m s$^{-1}$ (at 1.7 m) were
excluded, suggesting that at low wind speeds factors other than wind play a
more significant role in determining ROS.

\subsection{Fuel moisture content function}

\begin{figure}[ph!]
  \centering
  \includegraphics[width=16cm]{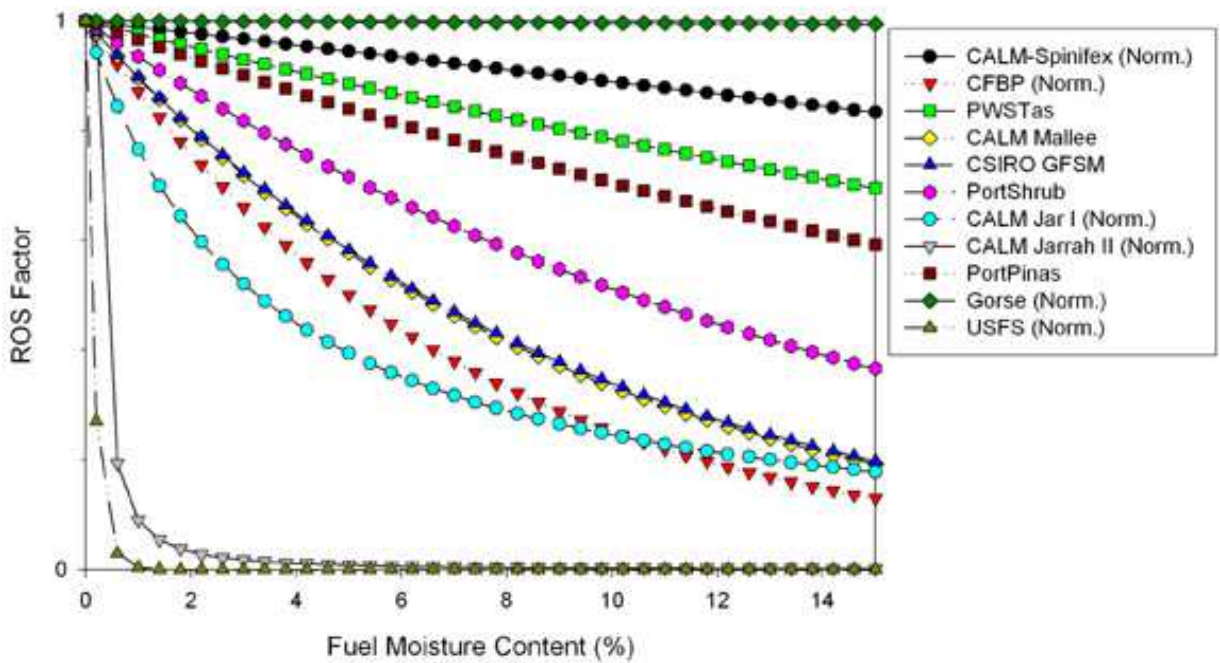}\\
  \caption{Graph of functional relationships between fuel moisture content and a rate of
  forward spread factor used in various empirical and quasi-empirical fire spread models.
  A number of models have been normalised in order to present them in conjunction with
  other models (i.e. norm.) The ROS factor shows the effect fuel moisture content has
  on the final rate of spread value in each model.}\label{Fig:FMCFn}
\end{figure}

Another method of comparing the various empirical and quasi-empirical models is
that of fuel moisture content function.  Not all the models discussed here
addressed the relation between fuel moisture content and rate of spread, and so
the discussion here is not comprehensive.  Figure \ref{Fig:FMCFn} shows the
various functions for those models that include fuel moisture content.

As with the wind function, there is a wide spread of functional forms used to
described the influence of FMC on ROS, perhaps reflecting modelling approaches,
methods or personal choice. There appear to be three types of functions
representing the fuel moisture content/ROS function: weakly linear (e.g. Gorse
(normalised) or CALM Spinifex (normalised)) in which the FMC plays a minor role
in determining the ROS, strongly exponential (e.g. CALM Jarrah II and USFS) in
which FMC plays a strong role until very low FMC values, and strongly linear or
weakly exponential (which might be approximated to linear) in which the role of
FMC is spread over a large range of values. The majority of models discussed
here fall into the latter group.

The weakest of the linear models (Gorse (normalised) is characterised by few
experiments with a limited range of FMC values, which raises the issue of how
many sample points are needed to properly inform functional choice and model
validation--one could argue that 9 fires is simply not enough given the range
of the uncontrolled variables.

The weakly exponential group of models, which includes strictly linear models
(e.g. CALM-Spinifex), appears to be the most robust in terms of range of FMC
values and experimentation.  It is interesting to note the similarity between
the CALM Mallee and the CSIRO Grass models.  The large difference in
functionality between the strongly exponential and weakly exponential is
interesting and may reflect differences in functionality as a result of wind
function modelling, as all modelling identified wind speed as the primary
variable and FMC as the secondary.

\subsection{Measurement issues}

All empirical science is limited by the ability to measure necessary
quantities, to quantify the errors in those measurements, and then relate those
measurements to the phenomenon under investigation and wildland fire science is
no different.  \citet{Sullivan2001} discussed the determination of the errors
in measuring wind speed under a forest canopy some distance from an
experimental fire and relating that measurement to measurements of fire spread.
The issues of where to measure (location, height, in the open, under the
canopy, etc.), how long to measure (instantaneous, period sampling, average,
period of average, etc.), and how to correlate measurements with observations
are complex and necessarily require approximations and simplification in order
to be undertaken.

Similarly, destructive sample of FMC has issues that complicate a seemingly
simple quantity.  The time of sampling (morning, afternoon, wetting period,
drying period), the general location (in the open or under the canopy, in the
sun, in the shade, in between, etc.), the specific location (surface litter
fuels, profile litter fuels, mid-layer fuels), the species of fuel (predominant
fuels, non-predominant fuel, live, dead, etc.). Also once the samples have been
taken there then is the issue of best drying methods for the particular samples
to ensure a water-free weight, the best method of determining an average value
for a plot, variance, error, etc.

Quantifying other factors such as fuel, again seemingly simple quantities rely
upon a knowledge of the mode of combustion of the fire and which aspects of the
fuel most influence that combustion and therefore the behaviour of the fire.
These include definition of fuel strata (which itself depends on the intensity
of the fire and which parts of the fuel complex will be burning during the fire
and thus contribute to the energy released by the fire), the structure of the
fuel and the size of fuel particles important to fire behaviour of the front,
flanks and behind the flame zone, the amount of fuel available, the amount
consumed, the chronology of the consumption of the fuel, the mode of
consumption, transport of burning fuel (i.e. firebrands), spatial and temporal
variation of these fuel and fire characteristics, determination of averages and
methods of averaging, determination of errors, etc.  The list could continue.
Other factors, such as air temperature and relative humidity, insolation,
atmospheric stability, slope, soil type and moisture, have their own range of
measurement difficulties, and are by no means the only quantities involved in
quantifying the behaviour of wildland fires.

Laboratory-based experiments may aim to reduce the variation and control the
errors in measurement of many of these quantities but are not immune to the
difficulties of measurement.

\subsection{Field versus laboratory experimentation}

Empirical or quasi-empirical modelling of fire behaviour has resulted in
significant advances in the state of wildfire science and produced effective
operational guides for determining the likely behaviour of wildfires for
suppression planning purposes.  Unlike physical or quasi-physical models of
fire behaviour, these systems are simple, utilise readily available fuel and
weather input data, and can be calculated rapidly.  However, there is a
significant difference between those models developed from field
experimentation and those developed from laboratory experimentation.

Large-scale field experiments are costly, difficult to organise, and inherently
have many of the difficulties associated with wildfire observations (e.g.
spatial and temporal variation of environmental variables, uncontrolled
variations, changing frames of reference and boundary conditions, etc.).
Laboratory experiments can be cheap and safe, provide relatively repeatable
conditions, and can limit the type and range of variations within variables and
thus simplify analysis. \citet{VanWagner1971} raises the issue of laboratory
versus field experimentation but avoids any categorical conclusions (perhaps
because there are none), simply stating some features of wildland fire
behaviour are better suited to studying in the controlled environment of the
laboratory or could not be attempted in the field, and other features cannot be
suitably replicated anywhere but in large-scale outdoor experiments.

Correct scaling of laboratory experiments (and field experiments for that
matter) is vital to replicating the conditions expected during a wildfire.
\citet{Byram1966} and \citet{Williams1969} conducted dimensional analysis of
(stationary) mass fires in order to develop scaling laws to conduct scaled
model experiments.  Both found that scaling across all variables presents
considerable difficulty and necessitates approximations, particularly in regard
to atmospheric variables, which result in impractical lower limits (e.g. model
forest fires $\simeq$ 6-16 m across \citep{Byram1966}, or gravitational
acceleration $\simeq$ 10 g \citep{Williams1969}). As a result, it is clear that
any scaled experiments must take great care in drawing conclusions that are
expected to be applicable at scales different from that of the experiments; not
only may physical and chemical processes behave differently at different scales
but the phenomena as a whole may behave differently.

The key difference between field-based and laboratory-based experimentation, in
this author's opinion, is the assumptions about the nature of combustion
(including heat transfer) that are required in order to design a useful
small-scale laboratory experiment. That is, there is the presumption implicit
in any laboratory experiment that there is sufficient understanding about the
nature of fire such that key variables can be isolated and measured without
regard to the fire itself.

One such aspect identified by \citet{Cheney1993} is the importance of the size
and shape of the fire in determining resultant fire behaviour.  Prior to this
work, it was thought that the size of the fire played little part in
determining the behaviour of a fire and thus the results of small experimental
fires could be extrapolated to larger fires burning under less mild conditions.
Other factors such as the physical structure of the fuel or moisture content of
live fuels, or other hitherto unconsidered factors, may play less significant
but important roles in explaining the unaccounted variation in ROS.

Field experiments on the other hand, by their very nature, are real fires and
thus incorporate all the interactions that define wildland fire. This aspect
holds considerable weight with end users who endow such systems with a
confidence that purely theoretical or laboratory-only-based models do not
receive. As \citet{Morvan2004a} concluded, no single approach to studying the
behaviour of wildland fire will provide a complete solution and thus it is
important that researchers maintain open and broad paradigm

\section{Acknowledgements}

I would like to acknowledge Ensis Bushfire Research and the CSIRO Centre for
Complex Systems Science for supporting this project; Jim Gould and Rowena Ball
for comments on the draft manuscript; and the members of Ensis Bushfire
Research who ably assisted in the refereeing process, namely Stuart Anderson,
Miguel Cruz, and Juanita Myers.

\bibliographystyle{apalike}

\begin{thebibliography}{}

\bibitem[Abbot and Burrows, 2003]{Abbott2003}
Abbot, I. and Burrows, N., editors (2003).
\newblock {\em Fire in Ecosystems of South-West Western Australia: Impacts and
  Management}.
\newblock Backhuys, Leiden, The Netherlands.

\bibitem[Alexander et~al., 1991]{Alexander1991}
Alexander, M., Stocks, B., and Lawson, B. (1991).
\newblock Fire behaviour in {Black Spruce}-lichen woodland: The {Porter Lake}
  project.
\newblock Information Report NOR-X-310, Forestry Canada, Northwest Region,
  Northern Forestry Centre, Edmonton, Alberta.

\bibitem[Andrews, 1986]{Andrews1986}
Andrews, P. (1986).
\newblock Behave: fire behaviour prediction and fuel modellings system - burn
  subsystem, part 1.
\newblock Technical Report General Technical Report INT-194, 130 pp., USDA
  Forest Service, Intermountain Forest and Range Experiment Station, Ogden, UT.

\bibitem[Baeza et~al., 2002]{Baeza2002}
Baeza, M., De~Lu\'is, M., Ravent\'os, J., and Escarr\'e, A. (2002).
\newblock Factors influencing fire behaviour in shrublands of different stand
  ages and the implications for using prescribed burning to reduce wildfire
  risk.
\newblock {\em Journal of Environmental Management}, 65(2):199--208.

\bibitem[Beer, 1991]{Beer1991}
Beer, T. (1991).
\newblock The interaction of wind and fire.
\newblock {\em Boundary-Layer Meteorology}, 54(2):287--308.

\bibitem[Beer, 1993a]{Beer1995}
Beer, T. (1993a).
\newblock Fire propagation in vertical stick arrays: {T}he effects of wind.
\newblock {\em International Journal of Wildland Fire}, 5(1):43--49.

\bibitem[Beer, 1993b]{Beer1993}
Beer, T. (1993b).
\newblock The speed of a fire front and its dependence on wind speed.
\newblock {\em International Journal of Wildland Fire}, 3(4):193--202.

\bibitem[Bilgili and Saglam, 2003]{Bigili2003}
Bilgili, E. and Saglam, B. (2003).
\newblock Fire behavior in maquis fuels in {T}urkey.
\newblock {\em Forest Ecology and Management}, 184(1-3):201--207.

\bibitem[Bradstock and Gill, 1993]{Bradstock1993}
Bradstock, R. and Gill, A. (1993).
\newblock Fire in semiarid, mallee shrublands - size of flames from discrete
  fuel arrays and their role in the spread of fire.
\newblock {\em International Journal of Wildland Fire}, 3(1):3--12.

\bibitem[Burgan, 1988]{Burgan1988}
Burgan, R. (1988).
\newblock 1988 revisions to the 1978 {National Fire-Danger Rating System}.
\newblock Research Paper SE-273, USDA Forest Service, Southeastern Forest
  Experiment Station, Asheville, North Carolina.

\bibitem[Burrows, 1994]{Burrows1994}
Burrows, N. (1994).
\newblock {\em Experimental development of a fire management model for jarrah
  ({\emph{Eucalyptus marginata} Donn ex Sm}) forest}.
\newblock PhD thesis, Dept of Forestry, Australian National University,
  Canberra.

\bibitem[Burrows, 1999a]{Burrows1999a}
Burrows, N. (1999a).
\newblock Fire behaviour in jarrah forest fuels: 1. {Laboratory experiments}.
\newblock {\em CALMScience}, 3(1):31--56.

\bibitem[Burrows, 1999b]{Burrows1999b}
Burrows, N. (1999b).
\newblock Fire behaviour in jarrah forest fuels: 2. {Field experiments}.
\newblock {\em CALMScience}, 3(1):57--84.

\bibitem[Burrows et~al., 1991]{Burrows1991}
Burrows, N., Ward, B., and Robinson, A. (1991).
\newblock Fire behaviour in spinifex fuels on the {Gibson Desert Nature
  Reserve, Western Australia}.
\newblock {\em Journal of Arid Environments}, 20:189--204.

\bibitem[Byram, 1966]{Byram1966}
Byram, G. (1966).
\newblock Scaling laws for modeling mass fires.
\newblock {\em Pyrodynamics}, 4:271--284.

\bibitem[Carrier et~al., 1991]{Carrier1991}
Carrier, G., Fendell, F., and Wolff, M. (1991).
\newblock Wind-aided firespread across arrays of discrete fuel elements.{ I.
  Theory}.
\newblock {\em Combustion Science and Technology}, 75:31--51.

\bibitem[Catchpole, 2000]{Catchpole2000}
Catchpole, W. (2000).
\newblock {\em FIRE! The Australian Experience. Proceedings of the 1999
  Seminar}, chapter The International Scene and Its Impact on Australia, pages
  137--148.
\newblock National Academies Forum.

\bibitem[Catchpole et~al., 1998a]{Catchpole1998b}
Catchpole, W., Bradstock, R., Choate, J., Fogarty, L., Gellie, N., McArthy, G.,
  McCaw, L., Marsden-Smedley, J., and Pearce, G. (1998a).
\newblock Co-operative development of equations for heathland fire behaviour.
\newblock In {\em Proceedings of III International Conference on Forest Fire
  Research, 14th Conference on Fire and Forest Meteorology, Luso, Portugal,
  16-20 November 1998, Vol 1}, pages 631--645.

\bibitem[Catchpole et~al., 1998b]{Catchpole1998a}
Catchpole, W., Catchpole, E., Butler, B., Rothermel, R., Morris, G., and
  Latham, D. (1998b).
\newblock Rate of spread of free-burning fires in woody fuels in a wind tunnel.
\newblock {\em Combustion Science and Technology}, 131:1--37.

\bibitem[Chandler et~al., 1983]{Chandler1983}
Chandler, C., Cheney, P., Thomas, P., Trabaud, L., and Williams, D. (1983).
\newblock {\em Fire in Forestry 1: Forest Fire Behaviour and Effects}.
\newblock John Wiley \& Sons, New York.

\bibitem[Cheney, 1981]{Cheney1981}
Cheney, N. (1981).
\newblock Fire behaviour.
\newblock In Gill, A., Groves, R., and Noble, I., editors, {\em Fire and the
  Australian Biota}, chapter~5, pages 151--175. Australian Academy of Science,
  Canberra.

\bibitem[Cheney and Gould, 1995]{Cheney1995}
Cheney, N. and Gould, J. (1995).
\newblock Fire growth in grassland fuels.
\newblock {\em International Journal of Wildland Fire}, 5:237--247.

\bibitem[Cheney et~al., 1993]{Cheney1993}
Cheney, N., Gould, J., and Catchpole, W. (1993).
\newblock The influence of fuel, weather and fire shape variables on
  fire-spread in grasslands.
\newblock {\em International Journal of Wildland Fire}, 3(1):31--44.

\bibitem[Cheney et~al., 1998]{Cheney1998}
Cheney, N., Gould, J., and Catchpole, W. (1998).
\newblock Prediction of fire spread in grasslands.
\newblock {\em International Journal of Wildland Fire}, 8(1):1--13.

\bibitem[Cheney et~al., 2001]{Cheney2001}
Cheney, P., Gould, J., and McCaw, L. (2001).
\newblock The dead-man zone-a neglected area of firefighter safety.
\newblock {\em Australian Forestry}, 64(1):45--50.

\bibitem[Cheney and Sullivan, 1997]{Cheney1997b}
Cheney, P. and Sullivan, A. (1997).
\newblock {\em Grassfires: Fuel, Weather and Fire Behaviour}.
\newblock CSIRO Publishing, Collingwood, Australia.

\bibitem[CSIRO, 1997]{CSIRO1997}
CSIRO (1997).
\newblock {CSIRO} {G}rassland {F}ire {S}pread {M}eter.
\newblock Cardboard meter.

\bibitem[Curry and Fons, 1940]{Curry1940}
Curry, J. and Fons, W. (1940).
\newblock Forest-fire behaviour studies.
\newblock {\em Mechanical Engineering}, 62:219--225.

\bibitem[Curry and Fons, 1938]{Curry1938}
Curry, J.~R. and Fons, W.~L. (1938).
\newblock Rate of spread of surface fires in the ponderosa pine type of
  california.
\newblock {\em Journal of Agricultural Research}, 57(4):239--267.

\bibitem[Deeming et~al., 1977]{Deeming1977}
Deeming, J., Burgan, R., and Cohen, J. (1977).
\newblock {The National Fire-Danger Rating System - 1978}.
\newblock Technical Report General Technical Report INT-39, USDA Forest
  Service, Intermountain Forest and Range Experiment Station, Ogden, UT.

\bibitem[Emmons, 1963]{Emmons1963}
Emmons, H. (1963).
\newblock Fire in the forest.
\newblock {\em Fire Research Abstracts and Reviews}, 5(3):163--178.

\bibitem[Emmons, 1966]{Emmons1966}
Emmons, H. (1966).
\newblock Fundamental problems of the free burning fire.
\newblock {\em Fire Research Abstracts and Reviews}, 8(1):1--17.

\bibitem[Fendell and Wolff, 2001]{Fendell2001}
Fendell, F. and Wolff, M. (2001).
\newblock {\em Wildland Fire Spread Models}, chapter 6: Wind-Aided Fire Spread,
  pages 171--223.
\newblock Academic Press, San Diego, CA, 1st edition.

\bibitem[Fernandes, 1998]{Fernandes1998}
Fernandes, P. (1998).
\newblock Fire spread modelling in {P}ortuguese shrubland.
\newblock In {\em Proceedings of III International Conference on Forest Fire
  Research, 14th Conference on Fire and Forest Meteorology, Luso, Portugal,
  16-20 November 1998, Vol 1}, volume~1, pages 611--628.

\bibitem[Fernandes, 2001]{Fernandes2001}
Fernandes, P. (2001).
\newblock Fire spread prediction in shrub fuels in {Portugal}.
\newblock {\em Forest Ecology and Management}, 144(1-3):67--74.

\bibitem[Fernandes et~al., 2002]{Fernandes2002}
Fernandes, P., Botelho, H., and Loureiro, C. (2002).
\newblock Models for the sustained ignition and behaviour of low-to-moderately
  intense fires in maritime pine stands.
\newblock page~98, Rotterdam, Netherlands. Millpress.
\newblock Proceedings of the IV International Conference on Forest Fire
  Research, Luso, Coimbra, Portugal 18-23 November 2002.

\bibitem[Fons, 1946]{Fons1946}
Fons, W.~L. (1946).
\newblock Analysis of fire spread in light forest fuels.
\newblock {\em Journal of Agricultural Research}, 72(3):93--121.

\bibitem[{Forestry Canada Fire Danger Group}, 1992]{FCFDG1992}
{Forestry Canada Fire Danger Group} (1992).
\newblock Development and structure of the {Canadian Forest Fire Behavior
  Prediction System}.
\newblock Information Report ST-X-3, Forestry Canada Science and Sustainable
  Development Directorate, Ottawa, ON.

\bibitem[Fransden, 1971]{Fransden1971}
Fransden, W. (1971).
\newblock Fire spread through porous fuels from the conservation of energy.
\newblock {\em Combustion and Flame}, 16:9--16.

\bibitem[Fransden, 1973]{Fransden1973}
Fransden, W.~H. (1973).
\newblock Using the effective heating number as a weighting factor in
  {R}othermel's fire spread model.
\newblock General Technical Report INT-10, USDA Forest Service, Intermountain
  Forest and Range Experiment Station, Ogden UT.

\bibitem[Gill et~al., 1995]{Gill1995}
Gill, A., Burrows, N., and Bradstock, R. (1995).
\newblock Fire modelling and fire weather in an {A}ustralian desert.
\newblock {\em CALMScience Supplement}, 4:29--34.

\bibitem[Gill et~al., 1981]{Gill1981}
Gill, A., Groves, R., and Noble, I., editors (1981).
\newblock {\em Fire and the Australian Biota}.
\newblock Australian Academy of Science, Canberra.

\bibitem[Gisborne, 1927]{Gisborne1927}
Gisborne, H. (1927).
\newblock The objectives of forest fire-weather research.
\newblock {\em Journal of Forestry}, 25(4):452--456.

\bibitem[Gisborne, 1929]{Gisborne1929}
Gisborne, H. (1929).
\newblock The complicated controls of fire behaviour.
\newblock {\em Journal of Forestry}, 27(3):311--312.

\bibitem[Goldammer and Jenkins, 1990]{Goldhammer1990}
Goldammer, J. and Jenkins, M., editors (1990).
\newblock {\em Fire in Ecosystem Dynamics}.
\newblock SPB Academic Publishing bv, The Hague, The Netherlands.

\bibitem[Grishin, 1984]{Grishin1984a}
Grishin, A. (1984).
\newblock Steady-state propagation of the front of a high-level forest fire.
\newblock {\em Soviet Physics Doklady}, 29(11):917--919.

\bibitem[Grishin, 1997]{Grishin1997}
Grishin, A. (1997).
\newblock {\em Mathematical modeling of forest fires and new methods of
  fighting them}.
\newblock Publishing House of Tomsk State University, Tomsk, Russia, english
  translation edition.
\newblock Translated from Russian by Marek Czuma, L Chikina and L Smokotina.

\bibitem[Hawley, 1926]{Hawley1926}
Hawley, L. (1926).
\newblock Theoretical considerations regarding factors which influence forest
  fires.
\newblock {\em Journal of Forestry}, 24(7):7.

\bibitem[Karplus, 1977]{Karplus1977a}
Karplus, W.~J. (1977).
\newblock The spectrum of mathematical modeling and systems simulation.
\newblock {\em Mathematics and Computers in Simulation}, 19(1):3--10.

\bibitem[Lawson et~al., 1985]{Lawson1985}
Lawson, B., Stocks, B., Alexander, M., and Van~Wagner, C. (1985).
\newblock A system for predicting fire behaviour in {C}anadian forests.
\newblock In {\em Eighth Conference on Fire and Forest Meteororology}, pages
  6--16.

\bibitem[Lee, 1972]{Lee1972}
Lee, S. (1972).
\newblock Fire research.
\newblock {\em Applied Mechanical Reviews}, 25(3):503--509.

\bibitem[Linn, 1997]{Linn1997}
Linn, R.~R. (1997).
\newblock A transport model for prediction of wildfire behaviour.
\newblock PhD Thesis LA-13334-T, Los Alamos National Laboratory.
\newblock Reissue of PhD Thesis accepted by Department of Mechanical
  Engineering, New Mexico State University.

\bibitem[Marsden-Smedley and Catchpole, 1995a]{Marsden1995a}
Marsden-Smedley, J. and Catchpole, W. (1995a).
\newblock Fire behaviour modelling in {Tasmanian} buttongrass moorlands {I.
  F}uel characteristics.
\newblock {\em International Journal of Wildland Fire}, 5(4):202--214.

\bibitem[Marsden-Smedley and Catchpole, 1995b]{Marsden1995b}
Marsden-Smedley, J. and Catchpole, W. (1995b).
\newblock Fire behaviour modelling in {Tasmanian} buttongrass moorlands {II.
  F}ire behaviour.
\newblock {\em International Journal of Wildland Fire}, 5(4):215--228.

\bibitem[McAlpine et~al., 1991]{McAlpine1991b}
McAlpine, R., Lawson, B., and Taylor, E. (1991).
\newblock Fire spread across a slope.
\newblock In {\em Proceedings of the 11th Conference on Fire and Forest
  Meteorology}, pages 218--225, Missoula, MT. Society of American Foresters.

\bibitem[McAlpine and Wakimoto, 1991]{McAlpine1991a}
McAlpine, R. and Wakimoto, R. (1991).
\newblock The acceleration of fire from point source to equilibrium spread.
\newblock {\em Forest Science}, 37(5):1314--1337.

\bibitem[McArthur, 1965]{McArthur1965}
McArthur, A. (1965).
\newblock Weather and grassland fire behaviour.
\newblock Country Fire Authority and Victorian Rural Brigades Association Group
  Officers Study Period, 13th - 15th August 1965.

\bibitem[McArthur, 1966]{McArthur1966}
McArthur, A. (1966).
\newblock Weather and grassland fire behaviour.
\newblock Technical Report Leaflet 100, Commonwealth Forestry and Timber
  Bureau, Canberra.

\bibitem[McArthur, 1967]{McArthur1967}
McArthur, A. (1967).
\newblock Fire behaviour in eucalypt forests.
\newblock Technical Report Leaflet 107, Commonwealth Forestry and Timber
  Bureau, Canberra.

\bibitem[McCaw, 1997]{McCaw1997}
McCaw, L. (1997).
\newblock {\em Predicting fire spread in {Western Australian} mallee-heath
  shrubland}.
\newblock PhD thesis, School of Mathematics and Statistics, University of New
  South Wales, Canberra, ACT, Australia.

\bibitem[Morvan et~al., 2004]{Morvan2004a}
Morvan, D., Larini, M., Dupuy, J., Fernandes, P., Miranda, A., Andre, J.,
  Sero-Guillaume, O., Calogine, D., and Cuinas, P. (2004).
\newblock Eufirelab: Behaviour modelling of wildland fires: a state of the art.
\newblock Deliverable D-03-01, EUFIRELAB.
\newblock 33 p.

\bibitem[Nelson, 2002]{Nelson2002}
Nelson, Jr., R. (2002).
\newblock An effective wind speed for models of fire spread.
\newblock {\em International Journal of Wildland Fire}, 11(2):153--161.

\bibitem[Nelson and Adkins, 1988]{Nelson1988}
Nelson, Jr., R.~M. and Adkins, C.~W. (1988).
\newblock A dimensionless correlation for the spread of wind-driven fires.
\newblock {\em Canadian Journal of Forest Research}, 18:391--397.

\bibitem[Pastor et~al., 2003]{Pastor2003}
Pastor, E., Zarate, L., Planas, E., and Arnaldos, J. (2003).
\newblock Mathematical models and calculation systems for the study of wildland
  fire behaviour.
\newblock {\em Progress in Energy and Combustion Science}, 29(2):139--153.

\bibitem[Peet, 1965]{Peet1965}
Peet, G. (1965).
\newblock A fire danger rating and controlled burning guide for the northern
  jarrah (euc. marginata sm.) forest of western australia.
\newblock Technical Report Bulletin No 74, Forests Department, Perth, Western
  Australia.

\bibitem[Perry, 1998]{Perry1998}
Perry, G. (1998).
\newblock Current approaches to modelling the spread of wildland fire: a
  review.
\newblock {\em Progress in Physical Geography}, 22(2):222--245.

\bibitem[Pyne et~al., 1996]{Pyne1996}
Pyne, S., Andrews, P., and Laven, R. (1996).
\newblock {\em Introduction to Wildland Fire, 2nd Edition}.
\newblock John Wiley and Sons, New York.

\bibitem[Pyne, 2001]{Pyne2001}
Pyne, S.~J. (2001).
\newblock {\em Year of the Fires : The Story of the Great Fires of 1910}.
\newblock Viking, New York.

\bibitem[Rothermel, 1972]{Rothermel1972}
Rothermel, R. (1972).
\newblock A mathematical model for predicting fire spread in wildland fuels.
\newblock Research Paper INT-115, USDA Forest Service.

\bibitem[Sauvagnargues-Lesage et~al., 2001]{Sauvag2001}
Sauvagnargues-Lesage, S., Dusserre, G., Robert, F., Dray, G., and Pearson, D.
  (2001).
\newblock Experimental validation in mediterranean shrub fuels of seven
  wildland fire rate of spread models.
\newblock {\em International Journal of Wildland Fire}, 10(1):15--22.

\bibitem[Sneeuwjagt and Peet, 1985]{Sneeuwjagt1985}
Sneeuwjagt, R. and Peet, G. (1985).
\newblock Forest fire behaviour tables for {Western Australia (3rd Ed.)}.
\newblock Department of Conservation and Land Management, Perth, WA.

\bibitem[Stocks et~al., 1991]{Stocks1991}
Stocks, B., Lawson, B., Alexander, M., Van~Wagner, C., McAlpine, R., Lynham,
  T., and Dub\'e, D. (1991).
\newblock The {Canadian} system of forest fire danger rating.
\newblock In Cheney, N. and Gill, A., editors, {\em Conference on Bushfire
  Modelling and Fire Danger Rating Systems}, pages 9--18, Canberra. CSIRO.

\bibitem[Stocks et~al., 2004]{Stocks2004}
Stocks, B.~J., Alexander, M.~E., and Lanoville, R.~A. (2004).
\newblock {Overview of the International Crown Fire Modelling Experiment
  (ICFME)}.
\newblock {\em Canadian Journal of Forest Research}, 34(8):1543--1547.

\bibitem[Sullivan, 2007]{Sullivan2007b}
Sullivan, A. (2007).
\newblock A review of wildland fire spread modelling, 1990-present, 1: Physical
  and quasi-physical models.
\newblock arXiv:0706.3074v1[physics.geo-ph], 46 pp.

\bibitem[Sullivan and Knight, 2001]{Sullivan2001}
Sullivan, A. and Knight, I. (2001).
\newblock Estimating error in wind speed measurements for experimental fires.
\newblock {\em Canadian Journal of Forest Research}, 31(3):401--409.

\bibitem[Taylor and Alexander, 2006]{Taylor2006}
Taylor, S. and Alexander, M. (2006).
\newblock Science, technology, and human factors in fire danger rating: the
  canadian experience.
\newblock {\em International Journal of Wildland Fire}, 15(1):121--135.

\bibitem[Thomas, 1967]{Thomas1967}
Thomas, P. (1967).
\newblock Some aspects of the growth and spread of fire in the open.
\newblock {\em Journal of Forestry}, 40:139--164.

\bibitem[Thomas and Pickard, 1961]{Thomas1961}
Thomas, P. and Pickard, R. (1961).
\newblock Fire spread in forest and heathland materials.
\newblock Report on forest research, Fire Research Station, Boreham Wood,
  Hertfordshire.

\bibitem[Van~Wagner, 1971]{VanWagner1971}
Van~Wagner, C. (1971).
\newblock Two solitudes in forest fire research.
\newblock Information Report PS-X-29, Canadian Forestry Service, Petawawa
  Forest Experiment Station, Chalk River, ON.

\bibitem[Van~Wagner, 1977a]{VanWagner1977b}
Van~Wagner, C. (1977a).
\newblock Conditions for the start and spread of crown fire.
\newblock {\em Canadian Journal of Forest Research}, 7(1):23--24.

\bibitem[Van~Wagner, 1977b]{VanWagner1977a}
Van~Wagner, C. (1977b).
\newblock Effect of slope on fire spread rate.
\newblock {\em Canadian Forestry Service Bi-Monthly Research Notes}, 33:7--8.

\bibitem[Van~Wagner, 1985]{vanWagner1985}
Van~Wagner, C. (1985).
\newblock Fire spread from a point source.
\newblock {Memo PI-4-20 dated January 14, 1985 to P. Kourtz (unpublished)},
  Canadian Forest Service, Petawawa National Forest Institute, Chalk River,
  Ontario.

\bibitem[Van~Wagner, 1987]{VanWagner1987}
Van~Wagner, C. (1987).
\newblock Development and structure of the canadian forest fire weather index
  system.
\newblock Forestry Technical Report~35, Canadian Forestry Service, Petawawa
  National.

\bibitem[Van~Wagner, 1988]{VanWagner1988}
Van~Wagner, C. (1988).
\newblock Effect of slope on fires spreading downhill.
\newblock {\em Canadian Journal of Forest Research}, 18:818--820.

\bibitem[Van~Wagner, 1998]{VanWagner1998}
Van~Wagner, C. (1998).
\newblock Modelling logic and the {Canadian Forest Fire Behavior Prediction
  System}.
\newblock {\em The Forestry Chronicle}, 74(1):50--52.

\bibitem[Viegas, 2002]{Viegas2002}
Viegas, D. (2002).
\newblock Fire line rotation as a mechanism for fire spread on a uniform slope.
\newblock {\em International Journal of Wildland Fire}, 11(1):11--23.

\bibitem[Viegas, 2006]{Viegas2006}
Viegas, D. (2006).
\newblock Parametric study of an eruptive fire behaviour model.
\newblock {\em International Journal of Wildland Fire}, 15(2):169--177.

\bibitem[Viegas et~al., 1998]{Viegas1998b}
Viegas, D., Ribeiro, P., and Maricato, L. (1998).
\newblock An empirical model for the spread of a fireline inclined in relation
  to the slope gradient or to wind direction.
\newblock In {\em III International Conference on Forest Fire Research. 14th
  Conference on Fire and Forest Meteorology Luso, Portugal, 16-20 November
  1998. Vol 1.}, pages 325--342.

\bibitem[Viegas, 1998]{Viegas1998a}
Viegas, D.~X. (1998).
\newblock Forest fire propagation.
\newblock {\em Philosophical Transmissions of the Royal Society of London A},
  356:2907--2928.

\bibitem[Viegas, 2005]{Viegas2005}
Viegas, D.~X. (2005).
\newblock A mathematical model for forest fires blowup.
\newblock {\em Combustion Science and Technology}, 177(1):27--51.

\bibitem[Weber, 1991]{Weber1991a}
Weber, R. (1991).
\newblock Modelling fire spread through fuel beds.
\newblock {\em Progress in Energy Combustion Science}, 17(1):67--82.

\bibitem[Weise and Biging, 1997]{Weise1997}
Weise, D.~R. and Biging, G.~S. (1997).
\newblock A qualitative comparison of fire spread models incorporating wind and
  slope effects.
\newblock {\em Forest Science}, 43(2):170--180.

\bibitem[Williams, 1969]{Williams1969}
Williams, F. (1969).
\newblock Scaling mass fires.
\newblock {\em Fire Research Abstracts and Reviews}, 11(1):1--23.

\bibitem[Williams, 1982]{Williams1982}
Williams, F. (1982).
\newblock Urban and wildland fire phenomenology.
\newblock {\em Progress in Energy Combustion Science}, 8:317--354.

\bibitem[Wolff et~al., 1991]{Wolff1991}
Wolff, M., Carrier, G., and Fendell, F. (1991).
\newblock Wind-aided firespread across arrays of discrete fuel elements. {II.
  E}xperiment.
\newblock {\em Combustion Science and Technology}, 77:261--289.

\end{thebibliography}


\newpage

\begin{table}[h!]
  \footnotesize
  \centering
  \caption{Summary of empirical models discussed in this paper}\label{Table:SummEmp}
  \begin{tabular}{cccccccc}
    \hline
    Model & Author & Year & Country & Field/Lab & Fuel type & No. fires & Size (w \ex l)  \\
          &        &      &           &           &      &     &  (m)\footnotemark \\
    \hline
    CFS-accel & McAlpine & 1991 & Canada & Lab & needles/Excel.& 29 & 0.915 \ex  6.15 \\
    CALM-Spin & Burrows  & 1991 & Aust. & Field & Spinifex   & 41 & 200 \ex 200\\
    CFBP      & FCFDG    & 1992 & Canada & Field & Forest    & 493 & 10-100 \ex 10-100\\
    Button   & Marsden-Smedley& 1995 & Aust.  & Field& Buttongrass  & 64 & 50-100\ex 50-100\\
    CALM Mallee & McCaw  & 1997 & Aust. & Field & Mallee/Heath & 18 & 200\ex 200\\
    CSIRO Grass& Cheney  & 1998 & Aust. & Field & Grass      & 121 & 100-200 \ex 100-300\\
    Heath & Catchpole & 1998 & Aust. & Field & heath/shrub   & 133 & 100\\
    PortShrub & Fernandes & 2001 & Portugal & Field & Heath/shrub & 29 & 10\\
    CALM Jarrah I & Burrows & 1999 & Aust. & Lab & Litter    & 144 & 2.0 \ex 4.0\\
    CALM Jarrah II & Burrows & 1999 & Aust. & Field & Forest & 56 & 100\\
    PortPinas & Fernandes & 2002 & Portugal & Field & Forest & 94 & 10-15\\
    Gorse & Baeza & 2002 & Spain & Field & gorse             & 9 & 33\\
    Maquis & Bilgili & 2003 & Turkey & Field & maquis        & 25 & 20\\
    CSIRO Forest & Gould & 2006 & Aust. & Field & Forest     & 99 & 200 \ex 200\\

    \hline
  \end{tabular}
\end{table}
\footnotetext{Where only one dimension is given by the authors, this is assumed
to be both width and length of the fire or plot}

\vspace{3cm}
\begin{table}[h!]
  \footnotesize
  \centering
  \caption{Summary of quasi-empirical models discussed in this paper}\label{Table:SummQE}
  \begin{tabular}{cccccccc}
    \hline
    Model & Author & Year & Country & Field/Lab & Fuel type & No. fires & Size (w \ex l)  \\
          &        &      &           &           &      &     &  (m)\\
    \hline
    TRW & Wolff & 1991 & USA & Lab & match splints & ? & 1.1 \ex  7 \\
    NBRU & Beer & 1993 & Aust. & Lab & match splints & 18 & 0.4 \ex 0.16 (2D)\\
    USFS  & Catchpole & 1998 & USA & Lab. & Pond./Excel & 357 & 1.0 \ex 8.0\\
    Coimbra & Viegas & 2002 & Spain & Lab & Pond. needles & 23 & 3.0 \ex 3.0\\
    Nelson & Nelson & 2002 & USA & Lab. & Birch sticks & 65 & ?\\

    \hline
  \end{tabular}
\end{table}

\begin{landscape}
\begin{table}
  \footnotesize
  \centering
  \caption{Summary of empirical models discussed in this paper}\label{Table:WindFn}
  \begin{tabular}{cccccccc}
    \hline
    Model & Field/Lab & Fuel type     & FMC Fn  & FMC    & Wind Fn     & Wind Range &   ROS Range\\
          &           &               &         & Range (\%) &         & (m s$^{-1}$) & (m s$^{-1}$)\\
    \hline
    \emph{Empirical}\\
    CFS-accel & Lab. & Pond./Excel    & -       & -      & -           & 0-2.22*     & ?-? \\
    CALM-Spinifex & Field & Spinifex  &$-82.08M$& 12-31  &  $U^2$      & 1.1-10      & 0-1.5 \\
    CFBP      & Field   & Forest     & e$^{-0.1386M}(1+M^{5.31})$ & ? &
    e$^{0.05039U}$ & ? & ?\\
    PWS-Tas   & Field& Buttongrass & e$^{-0.0243M}$&8.2-96&$U^{1.312}$& 0.2-10      & ?-?\\
    CALM Mallee & Field & Mallee   & e$^{-0.11M_{ld}}$&4-32&$U^{1.05}$& 1.5-6.9     & 0.13-6.8\\
    CSIRO Grass & Field & Grass & e$^{-0.108M}$ & 2.7-12.1 &$U^{0.844}$& 2.9-7.1    & 0.29-2.07\\
    Heath & Field & heath/shrub & NA            & NA     &$U^{1.21}$  &0.11-10.1    & 0.01-1.00\\
    PortShrub & Field & Heath/shrub&e$^{-0.067M}$&10-40  &e$^{0.092U}$&0.28-7.5     & 0.01-0.33\\
    CALM Jarrah I & Lab.&Litter&$\frac{1}{0.003+0.000922M}$&3-14&$U^{2.22}$ & 0.0-2.1 & 0.002-0.075\\
    CALM Jarrah II & Field & Forest&23.192$M^{-1.495}$&3-18.6&$U^{2.674}$ & 0.72-3.33 & 0.003-0.28\\
    PortPinas & Field & Forest& e$^{-0.035M}$&8-56  &$U^{0.868}$ &
    0.3-6.4&0.004-0.231\\

    Gorse & Field     & gorse         & -0.0004M& 22-85  & NA          &$<$ 1.4     & 0.004-0.039\\
    Maquis & Field    & maquis        & NA      & 15.3-27.7&$0.495U$   & 0.02-0.25  & 0.01-0.15\\
    \ignore{CSIRO Forest & Field & Forest     & ?       & ?      & ?           & ?-?        & ?-?\\
    \emph{Quasi-empirical}\\}
    TRW & Lab. & match splints        & NA      & NA     & $U^{0.5}$   & 0-4.7      & 0-0.007\\
    NBRU & Lab. & match splints       & NA      & NA     & $U^3$       & 0-9        & 0.004-0.38\\
    USFS & Lab. & Pond./Excel &$\frac{\exp^{-4.05M}}{(700+2260M)}$& 2-33 &$U^{0.91}$ & 0-3.1   & 0-0.23\\
    Coimbra & Lab.& Pond. needles   & NA      & 10-15  & - & ? & ?\\
    Nelson & Lab./Field & Birch sticks& NA     & NA      & $U^{1.51}$  & 0.0-3.66   & $<$ 0.271\\

    \hline
  \end{tabular}

\end{table}
\end{landscape}

\end{document}